\documentclass[aps,prd,superscriptaddress,showpacs,floatfix,11pt]{revtex4-1}

\usepackage{color}
\usepackage{ulem}
\usepackage{hyperref}

\def \be {\begin{equation}}
\def \ee {\end{equation}}
\def \ba {\begin{array}}
\def \ea {\end{array}}
\def \bea {\begin{eqnarray}}
\def \eea {\end{eqnarray}}

\def \ble {\begin{widetext}\begin{equation}}
\def \ele {\end{equation}\end{widetext}}
\def \blea {\begin{widetext}\begin{eqnarray}}
\def \elea {\end{eqnarray}\end{widetext}}

\def \nn {\nonumber}

\newcommand{\eq}[1]{(\ref{#1})}

\def \a {\alpha}
\def \b {\beta}
\def \g {\gamma}

\def \d {\delta}

\def \e {\epsilon}

\def \r {\rho}

\def \mA {\mathcal A}
\def \mB {\mathcal B}

\def \mD {\mathcal D}
\def \mE {\mathcal E}

\def \mH {\mathcal H}
\def \mI {\mathcal I}

\def \mX {\mathcal X}
\def \mY {\mathcal Y}

\def \p {\partial}
\def \f {\frac}

\def \mc {\mathcal}

\def \td {\tilde}

\def \inf {\infty}

\def \lag {\langle}
\def \rag {\rangle}

\def \ep {\mathrm{e}}
\def \ii {\mathrm{i}}

\def \and {{\textrm{and}}}
\def \with {{\textrm{with}}}

\def \CFT {{\textrm{CFT}}}

\def \ET {{\langle T(w)\rangle_\rho}}
\def \EA {{\langle \mathcal{A}(w)\rangle_\rho}}
\def \EB {{\langle \mathcal{B}(w)\rangle_\rho}}
\def \ED {{\langle \mathcal{D}(w)\rangle_\rho}}
\def \EE {{\langle \mathcal{E}(w)\rangle_\rho}}
\def \EH {{\langle \mathcal{H}(w)\rangle_\rho}}
\def \EI {{\langle \mathcal{I}(w)\rangle_\rho}}

\def \ET {\langle T\rangle_\rho}
\def \EA {\langle \mathcal{A}\rangle_\rho}
\def \EB {\langle \mathcal{B}\rangle_\rho}
\def \ED {\langle \mathcal{D}\rangle_\rho}
\def \EE {\langle \mathcal{E}\rangle_\rho}
\def \EH {\langle \mathcal{H}\rangle_\rho}
\def \EI {\langle \mathcal{I}\rangle_\rho}
\def \pET {\partial\langle T\rangle_\r}
\def \ppET {\partial^2\langle T\rangle_\r}
\def \pppET {\partial^3\langle T\rangle_\r}
\def \pfourET {\partial^4\langle T\rangle_\r}

\def \pEA {\partial\langle \mathcal{A}\rangle_\r}

\begin{document}

\title{Non-geometric States in a Holographic Conformal Field Theory}
\hfill{NCTS-TH/1808}
\author{Wu-zhong Guo}
\email{wzguo@cts.nthu.edu.tw}
\affiliation{Physics Division, National Center for Theoretical Sciences, National Tsing Hua University,\\
No.\ 101, Sec.\ 2, Kuang Fu Road, Hsinchu 30013, Taiwan}

\author{Feng-Li Lin}
\email{linfengli@phy.ntnu.edu.tw}
\affiliation{Department of Physics, National Taiwan Normal University,
No.\ 88, Sec.\ 4, Ting-Chou Road, Taipei 11677, Taiwan}

\author{Jiaju Zhang}
\email{jiaju.zhang@unimib.it}
\affiliation{Dipartimento di Fisica G.\ Occhialini, Universit\`a degli Studi di Milano-Bicocca,\\Piazza della Scienza 3, 20126 Milano, Italy}
\affiliation{INFN, Sezione di Milano-Bicocca, Piazza della Scienza 3, 20126 Milano, Italy}

\begin{abstract}
  In the AdS$_3$/CFT$_2$ correspondence, we find some conformal field theory (CFT) states that have no bulk description by the Ba\~nados geometry. We elaborate the constraints for a CFT state to be geometric, i.e., having a dual Ba\~nados metric, by comparing the order of central charge of the entanglement/R\'enyi entropy obtained respectively from the holographic method and the replica trick in CFT. We find that the geometric CFT states fulfill Bohr's correspondence principle by reducing the quantum KdV hierarchy to its classical counterpart. We call the CFT states that satisfy the geometric constraints geometric states, and otherwise non-geometric states. We give examples of both the geometric and non-geometric states, with the latter case including the superposition states and descendant states.
\end{abstract}

\maketitle


\section{Introduction}


   The anti-de Sitter/conformal field theory (AdS/CFT) correspondence {conjectures} that the bulk quantum gravity is equivalent to the boundary CFT \cite{Maldacena:1997re}. In the semi-classical limit of bulk theory, a CFT state is believed to be dual to a bulk geometry if the quantum fluctuation can be minimized. We call such kind of CFT states the geometric states. Thus, it is easy to see that the superposition of two geometric states cannot be geometric because the superposition principle should not hold for the bulk classical gravity  \cite{Almheiri:2016blp}. Despite there are many discussions on the criterion for a CFT state to be geometric, e.g., \cite{Beach:2016ocq,Balasubramanian:2007zt} and a review in \cite{Rangamani:2016dms}, it {still lacks} a concise criterion that one can adopt to check for more generic cases. For example, in AdS$_5$/CFT$_4$ correspondence {people} know that the vacuum state of ${\rm SU}(N)$ gauge theory admits only planar correlators in the large $N$ limit, which is then dual to classical gravity in pure AdS$_5$ space.  In this case, the quantum fluctuation of non-planar diagrams are suppressed and a bulk geometry is emerging as the holographic dual. However, there is no clear planar limit for arbitrary excited states.

The situation becomes sharper in three-dimensional (3D) AdS gravity which is dual to a two-dimensional (2D) CFT \cite{Brown:1986nw}, and thus the bulk Ba\~nados geometries \cite{Banados:1998gg} are determined by the expectation value of stress tensor of dual 2D CFT states in the large central charge $c$ limit. Due to the topological nature of 3D AdS gravity, we can state that the 2D geometric states should be described by the Ba\~nados geometries.
The primary states and canonical ensemble states are known to be described by the Ba\~nados geometries as can be verified by the match of entanglement entropy and its holographic dual \cite{Ryu:2006bv,Hubeny:2007xt} in the Ba\~nados-Teitelboim-Zanelli black hole \cite{Banados:1992wn} background.
Here $c$ plays the similar role as $N$ in the AdS$_5$/CFT$_4$, however, there is no analogue of planar limit even for vacuum state to define the suppression of quantum fluctuation. Naively, one can require the standard deviation/uncertainty of any local operator to be small as the criterion for the suppression of quantum fluctuation, and thus the geometric states. However, the question is what is the exact suppression order of these standard deviations/uncertainties should be in the large $c$ expansion. We need a concise criterion to check for more generic (non-)geometric states, at least in AdS$_3$/CFT$_2$.

  In this work, we formulate such a criterion by comparing the non-local observables such as entanglement entropy and R\'enyi entropy with their holographic duals \cite{Ryu:2006bv,Hubeny:2007xt,Dong:2016fnf}. If the CFT state is geometric, then its entanglement/R\'enyi entropy calculated \'a la replica trick \cite{Chen:2016lbu,Lin:2016dxa,He:2017vyf,He:2017txy} should agree with the corresponding holographic dual calculated from the dual Ba\~nados geometries. Otherwise, it is non-geometric. Moreover, by short interval expansion, we can turn this criterion into the constraints on the standard deviation of the stress tensors and its higher order cousins in terms of Korteweg-de Vries (KdV) charges. This will then tell precisely how much the quantum fluctuation should be suppressed for a state to be geometric.
  With such a concrete {criterion}, we indeed find some new non-geometric states, which are descendant states.

    Our paper is organized as follows.
    In section~\ref{sec2} we state explicitly our criterion for the geometric CFT states.
    In section~\ref{sec3}, we derive the conditions for geometric CFT states on the expectation values of quasi-primaries.
    In section~\ref{sec4}, we demonstrate a correspondence principle for the KdV charges for the geometric CFT states.
    We then give the examples for the geometric CFT states and non-geometric CFT states in sections~\ref{sec5} and \ref{sec6}, respectively.
    Finally, we conclude our paper in section~\ref{sec7} with discussions on our geometric state conditions and the connected correlation functions characterizing the suppression of quantum fluctuations.
    Besides, we elaborate technical details in various appendices.
    In Appendix~\ref{appA}, we give the more explicit details of the conditions given in section~\ref{sec3} for geometric CFT states.
    In Appendix~\ref{appB}, we elaborate the derivation of the conditions in section~\ref{sec3} and Appendix~\ref{appA}.
    In Appendix~\ref{appC}, we give the detail check for a coordinate-dependent example of geometric state discussed in section~\ref{sec5}.
    In Appendix~\ref{appD}, we elaborate the check of non-geometric descendant states discussed in section~\ref{sec6}.

\section{Criterion for geometric CFT states in Ba\~nados geometry} \label{sec2}

Due to the topological nature of 3D Einstein gravity, i.e., {that there is} no bulk propagating degree of freedom, the bulk geometry is completely determined  by the asymptotic boundary constraints, this then leads Ba\~nados to conjecture that all the vacuum asymptotically AdS$_3$ solutions of 3D Einstein gravity are completely classified by the boundary conformal symmetries.  Applying this conjecture to AdS/CFT correspondence, it leads to the Ba\~nados geometries which are determined by the expectation value of stress tensor with respect to the dual CFT state. More precisely, the form of the Ba\~nados geometry takes the form \cite{Banados:1998gg}
\be\label{Banados}
ds^2= \frac{dy^2}{y^2}
     +\frac{L_\r}{2} dz^2
     +\frac{\bar L_\r} {2}d\bar z^2
     +\Big(\frac{1}{y^2}+\frac{y^2}{4}L_\r{\bar L_\r}\Big) dz d\bar z,
\ee
where we set the AdS radius to unity $R=1$ so that the bulk Newton constant $G_N$ is related to the central charge $c$ of the dual CFT by $c=\frac{3}{2G_N}$ \cite{Brown:1986nw}.

We consider a holographic CFT on a cylinder with complex coordinate $w$ and spatial period $L$ in a state with density matrix $\r$, and the cylinder can be mapped to a complex plane with coordinate $z$ by the conformal transformation $z=\ep^{\f{2\pi\ii w}{L}}$.
The functions $L_{\rho}(z)$, $\bar L_{\rho}(\bar z)$ in the Ba\~nados geometry are respectively holomorphic and anti-holomorphic, and are related to expectation value of stress tensor on the plane with respect to the dual CFT state
\be
\langle T(z) \rangle_{\rho}=-\frac{c}{12}L_{\rho}(z), ~~ \langle \bar T (\bar z) \rangle_{\rho}=-\frac{c}{12}\bar L_{\rho}(\bar z).
\ee

Given a Ba\~nados geometry which is dual to a CFT state $\r$, one can then evaluate the holographic entanglement/R\'enyi entropy \'a la the prescriptions in \cite{Ryu:2006bv,Hubeny:2007xt,Dong:2016fnf}.  Both the holographic entanglement and R\'enyi entropies are given by the area law formula.  If we consider a CFT state, for which $\langle T(z)\rangle_\rho$ and $\langle \bar T(z)\rangle_\rho $ are of order $c$, then the metric of the dual Ba\~nados geometry is of order $c^0$ in the large $c$ expansion, and should be independent of $c$ in the large $c$ limit. Thereby, the area of minimal surface or cosmic brane should be independent of $c$ so that the holographic entanglement/R\'enyi entropies should be of order $c$ due to the relation $c={3\over 2G_N}$. Based on the above result, we now formulate our criterion for the geometric CFT states:

\textit{For a 2D CFT state of order $c$ stress tensor expectation value to be holographic dual to a Ba\~nados geometry, the entanglement/R\'enyi entropy obtained from CFT calculations should be at most order $c$ in the large $c$ limit. Otherwise, we call the CFT state non-geometric.}

\section{Constraints for geometric CFT states}\label{sec3}

Based on our proposed criterion for the geometric CFT states, i.e., that the entanglement/R\'enyi entropy should be at most of order $c$ in the large $c$ limit, we would like to extract the necessary constraints by explicitly evaluating the entanglement/R\`enyi entropy.  The prescription of evaluating entanglement/R\'enyi entropy is based on the replica trick \cite{Calabrese:2004eu}, which leads to an $n$-fold CFT that we call $\CFT^n$.
However, there is usually no closed form of entanglement/R\'enyi entropy for generic excited states. Instead we will evaluate in the short-interval expansion, similar to what has done in \cite{Chen:2016lbu,Lin:2016dxa,He:2017vyf,He:2017txy}. By assuming dominance of the vacuum conformal family in the operator product expansion (OPE) of twist operators \cite{Headrick:2010zt,Calabrese:2010he,Rajabpour:2011pt,Chen:2013kpa} in the large $c$ limit, the entanglement/R\'enyi entropy takes the formal form in terms of the series of expectation values of $\CFT^n$ quasiprimary fields $\Phi_K$ that are constructed by operators in the vacuum conformal family of the original one-fold CFT. Since the contributions from the holomorphic and anti-holomorphic sectors decouple and are similar, in this paper we only consider the contributions from the holomorphic sector.

We consider the short interval $A=[w,w+\ell]$ with $\ell\ll L$, and from OPE of twist operators we get the short interval expansion of the R\'enyi entropy
\bea \label{eq3}
&& S_{A,\rho}^{(n)} = \f{c(n+1)}{12n}\log\f{\ell}{\e}\\
&& \phantom{S_{A,\rho}^{(n)} =}
                    - \f{1}{n-1}\log \Big( \sum_K d_K \sum_{r=0}^\inf \f{a_K^r}{r!} \ell^{h_K+r} \lag \Phi_K^{(r)}(w) \rag_\r \Big).\nn
\eea
The summation of $K$ is over all the $\CFT^n$ holomorphic quasiprimary operators $\Phi_K$, with conformal weight $h_K$, which are constructed from the holomorphic quasiprimary operators in the original one-fold CFT.
The forms of $\Phi_K$ to level 8, which are constructed from $T$ at level 2, $\mA$ at level 4, $\mB$, $\mD$ at level 6, and $\mE$, $\mH$, $\mI$ at level 8, as well as their corresponding OPE coefficients $d_K$, can be found in \cite{Chen:2013dxa}.
There is the coefficient $a_K^r={C_{h_K+r-1}^r}/{C_{2h_K+r-1}^r}$.

Requiring that the R\'enyi entropy of $A$ in state $\r$ is of at most order $c$, we get the constraints for the one-point functions up to level $6$,
\bea \label{constraints}
&& \ET = c \alpha (w)+\beta (w)+\frac{\gamma (w)}{c}+O\Big(\f1{c^2}\Big), \nn\\
&& \EA = c^2 \alpha (w)^2+c \delta (w)+\epsilon (w)+O\Big(\frac{1}{c} \Big), \nn\\
&& \EB = c^2 \Big[\alpha'(w)^2-\frac{4}{5} \alpha (w) \alpha''(w)\Big]+ O(c), \\
&& \ED = c^3 \alpha (w)^3+3 c^2 \alpha (w) [\delta (w)-\alpha (w) \beta (w)]
         +O(c), \nn
\eea
with $\a(w)$, $\b(w)$, $\g(w)$, $\d(w)$, $\e(w)$ being arbitrary order $O(c^0)$ holomorphic functions.

    We  write the conditions (\ref{constraints}) as the suggestive forms
\bea
&&\label{Tsigma} \lim_{c\to \infty}\frac{\langle \mathcal{A}\rangle_\rho -\langle T\rangle^2_\rho}{c^2}=0\;,\\
&& \label{Tcubic} \lim_{c\to \infty}\frac{\langle \mathcal{D}\rangle_\rho-3 \langle \mathcal{A}\rangle_\rho \langle T\rangle_\rho +2 \langle T\rangle_\rho^3}{c^2}=0.
\eea
Recall that $\mA = (TT) - \f{3}{10}\p^2 T$ with $(\cdots)$ denoting the normal ordering,  $\sqrt{\langle \mathcal{A}\rangle_\rho -\langle T\rangle^2_\rho}$ plays the role of standard deviation of $T$ with respect to the geometric state $\rho$ and thus \eq{Tsigma}  tells that it should be smaller than order $c$ in the large $c$ limit. Similarly, $\mD= (T(TT))+O(T^2)$, thus \eq{Tcubic} suggests that the uncertainty of cubic quantum fluctuation of $T$ should be also not larger than order $c$. There are more constraints at higher orders of $\ell$.
See Appendix~\ref{appA} and \ref{appB} for more details.

   Note that these constraints are in analogy to the planar limit of the large $N$ expansion in 4D Yang-Mills theory for the vacuum state. However, we are considering excited state of large $c$ 2D CFTs, and there is no known planar limit for this case. Instead, our simple criterion serves as a guide for the analogy quantum suppression, and yields the precise constraints for the geometric states. Next we will justify the semi-classical nature of the geometric states  for the physical observables in the sense of Bohr's correspondence principle.

\section{Quantum to classical KdV equation and charges for geometric CFT states}\label{sec4}

 The geometric state constraints relate the expectation values of operators in the vacuum family quasiprimaries. We will show that these constraints in fact reduce the quantum KdV equation and charges to their classical counterparts.

For demonstration, we write down the quantum KdV currents up to level 6 \cite{Sasaki:1987mm,Eguchi:1989hs,Kupershmidt:1989bf}
\be
J_2 = T, ~~
J_4 = (TT), ~~
J_6 = (T(TT)) - \f{c+2}{12}(T'T'),
\ee
with the parenthesises $(\cdots)$ denoting the normal ordering operators.
In terms of the quasiprimary operators and their derivatives, we obtain
\bea
&& J_2 = T, ~~
   J_4 = \mA + \f{3}{10} T'', \nn\\
&& J_6 = \mD
         -\frac{25 (2 c+7) (7 c+68)}{108 (70 c+29)} \mB \nn\\
&& \phantom{J_6=}
         - \frac{2 c - 23}{108} \mA''
         - \frac{c-14}{280} T^{(4)} ,
\eea
These currents form the mutually commuting KdV charges
\be
Q_{2k-1} = \int_0^L \f{dw}{L} J_{2k}(w),
\ee
which constitute the integrability hierarchy of the quantum KdV equation
\be
\dot T = \f{1-c}{6}T''' - 3(TT)'
       = - \f{5c+22}{30} T''' - 3 \mA'.
\ee
Using the leading order geometric state constraints (\ref{constraints}), we set $\a(w)= U(w)/6$ and get the classical KdV equation
\be \label{clKdVeq}
\dot U = U'''+6 U U'.
\ee
Note that $\p_t$, which we denote by dot, has been rescaled from the quantum KdV equation to its classical counterpart.

In the large $c$ limit, a natural definition of the classical counterpart of quantum KdV currents with respect to state $\r$ is
\be
J^{\rho}_{2k}(w)\equiv \lim_{c\to \infty}\frac{6^{k}}{c^{k}}\langle J_{2k}(w)\rangle_{\rho}.
\ee
Using the leading order of (\ref{constraints}) we can then turn $J^{\rho}_{2k}$ into the standard classical form
\bea \label{Jrho2468}
 J^{\rho}_{2} = U, ~~
   J^{\rho}_{4} = U^2, ~~
   J^{\rho}_{6} = U^3-\frac{1}{2}U'^2\;.
\eea
Their associated KdV charges constitute the integrability hierarchy of the classical KdV equation (\ref{clKdVeq}). This reflects Bohr's correspondence principle for these geometric states by reducing these KdV conserved currents into their classical counterparts.

 In the textbook \cite{DiFrancesco:1997nk}, the quantum to classical reduction for the KdV equation is obtained by simply replacing the KdV current operator with its classical counterpart without referring to the associated state.
 This does not work if the associated CFT state is non-geometric, as we discuss in this paper.

\section{Examples of geometric CFT states}\label{sec5}

In \cite{Asplund:2014coa,Caputa:2014eta,Fitzpatrick:2015zha,Hijano:2015qja}, it has been shown that the R\'enyi entropy in the primary excited state
\be
\r_\phi = \f{1}{\a_\phi}|\phi\rag\lag\phi|,
\ee
is of order $c$ if the conformal weight $h_{\phi}$ is at most of order $c$, so that they should satisfy all the geometric state constraints \eq{constraints}. 
This is also consistent with the calculation \cite{Lin:2016dxa,He:2017vyf} from OPE of twist operators to order $\ell^8$.

Even without an explicit check as done for the primary states, we can argue that some particular states should satisfy the geometric state constraints.  For example, the thermal states which are dual to BTZ black holes, thus should also be geometric. Similarly, the states which are conformally related to the vacuum state on the plane, denoted by $|0\rangle$ should also be geometric. In the bulk, these states are dual to the Ba\~nados geometries which can be transformed to pure AdS$_3$ by the coordinate transformation dual to the boundary conformal map. These states include the thermal state and  the conical defect state.

In quantum mechanics the wave-packet state behaves like a classical particle. This motivates us now to check if a wave-packet state can also have the bulk description. Explicitly, the state considered has the density matrix
\bea
&& \r_{\phi(w_0)} = \f{1}{\a_\phi}\Big[ \f{L}{\pi} \sin\f{\pi(\bar w_0 - w_0)}{L} \Big]^{2h_\phi}
                  \phi(w_0)|0\rag\lag0|\phi(\bar w_0) \nn\\
&& \phantom{\r_{\phi(w_0)}}
                = \f{1}{\a_\phi} \Big( \f{1-z_0\bar z_0}{\bar z_0} \Big)^{2h_\phi} \phi(z_0)|0\rag\lag0|\phi(1/\bar z_0).
\eea
Note that $w_0$ is a position on the cylinder and $z_0$ is a position on the plane with the relation $z_0=\ep^{\f{2\pi\ii w_0}{L}}$.
Since $\phi(z_0)|0\rag = \ep^{z_0 L_{-1}} |\phi\rag$, the above state can be understood as a coherent sum of the primary state $|\phi\rangle$ and its global descendants.
We check that the one point functions in the state $\r_{\phi(w_0)}$ satisfy the constraints (\ref{constraints}). See Appendix~\ref{appC} for details.
This is consistent with the fact that on the cylinder the locally excited state is dual to a moving particle in AdS$_3$ \cite{Nozaki:2013wia,Asplund:2014coa}, i.e., that there exists a bulk geometric description.

\section{Examples of non-geometric CFT states}\label{sec6}

From our discussions we see that there are an infinite tower of constraints for a state to be geometric. Then it seems that it should be quite easy to have non-geometric states by violating one of the infinite number of constraints. The reason why we did not know any example of non-geometric states is partly due to lack of principle of check as proposed in this work, and partly due to the technical involvement of evaluating the geometric state constraints.
In the following we will consider some examples of non-geometric states, for which we know how to evaluate the associated one-point functions of the vacuum family quasiprimary operators to check \eq{constraints}.

As discussed in the introduction, one expects the superposition of primary states will not be geometric because the bulk gravity is classical so that the superposition principle does not work. Now we would like to check this explicitly.

Let us choose $|\phi_1\rag$ and $|\phi_2\rag$ as two primary states with conformal weights $h_{\phi_1}=c\e_{\phi_1} + O(c^0)$,  $h_{\phi_2}=c\e_{\phi_2} + O(c^0)$, and $\e_{\phi_1} \neq \e_{\phi_2}$.
We consider the superposition state
\be \label{superposition}
\cos(\theta) |\phi_1\rangle+\ep^{\ii\psi}\sin(\theta)|\phi_2\rangle.
\ee
The constraints (\ref{constraints}) are satisfied separately for the states $|\phi_1\rangle$ and $|\phi_2\rangle$, however they are violated for the superposition state (\ref{superposition}).
This means that the superposition of two primary states is non-geometric as we expect.
It is straightforward to generalize the above result to superposition states  $\sum_i c_i |\phi_i\rangle$ with $|\phi_i\rangle$'s being different primary states.

Other examples that do not satisfy the constraints (\ref{constraints}) are some descendant states
\bea
&& |\phi^{(m)}\rag ~ \with ~ h_\phi+m \sim O(c) , \nn\\
&& |\td\phi\rag ~ \with ~ h_\phi \sim O(c), \nn\\
&& |\td\phi^{(m)}\rag ~ \with ~ h_\phi+m \sim O(c) , \nn\\
&& |T^{(m)}\rag ~ \with ~ m \sim O(c) , \nn\\
&& |\mA^{(m)} \rag ~ \with ~ m \sim O(c),
\eea
where $\phi$ is a primary operator and $\td\phi$ is a quasiprimary operator with the definition $\td\phi \equiv (T\phi)-\frac{3}{2(h_\phi+1)}\phi''$. Note that we have not yet normalize these descendant states properly.
By $h_\phi+m \sim O(c)$, we mean that either $h_\phi$ or $m$ can be of order $O(c^0)$ or $O(c)$ but the sum $h_\phi+m$ is of order $O(c)$. See more details in Appendix~\ref{appD}.


Among the examples of non-geometric states, the superposition states can be understood intuitively. On the other hand, we have no immediate understanding why the descendant states lack the bulk classical geometric descriptions. In \cite{Fitzpatrick:2015foa}, the descendant states are understood as the dressings of gravitons on the particle's worldline. It is hard to see why some of the dressings cannot be backreacted geometrically, especially for the case with $m$ being $O(c^0)$. We may then ask if these states will turn to be geometric if quantum gravity effects are taken into account. In the context of perturbative quantum gravity by including higher derivative curvature terms, the answer is no because these terms are of higher orders in $G_N\sim 1/c$ so that they can only yield subleading order $1/c$ corrections to the Ba\~nados geometry, and the holographic entanglement/R\'enyi entropies remain order $c$. Therefore, we are forced to accept the existence of these non-geometric states, or the quantum gravity correction should be non-perturbative.

Moreover, in the context of quantum thermalization and canonical typicality \cite{Goldstein:2005aib,Popescu:2005} the non-geometric states are obviously the atypical states because their entanglement/R\'enyi entropies are quite different from the ones of thermal states. Using the result in \cite{Kraus:2016nwo}, it can be shown that there are more descendant states than the primary ones at high levels in the large $c$ limit \cite{Guo:2018pvi}. If most of these descendant states are non-geometric, one would then expect the canonical typicality to fail for 2D large $c$ CFTs.

\section{Discussions: Geometric state conditions and connected correlation functions}\label{sec7}

  In this work, based on (holographic) entanglement entropy we have formulated a criterion to check if a 2D CFT state can have a bulk geometric description or not. Moreover, we derive the explicit constraints for explicit check, and find that all the primary states are geometric along with the discovery of some non-geometric states.

  In this concluding section, we elaborate the relation between our geometric state conditions and the connected correlation functions which characterize the suppression of the quantum fluctuations.

In statistical mechanics the connected correlation function or Ursell function of multivariate random variables  is defined by
\bea
U_n(X_1,X_2,...,X_n):=\f{\partial}{\partial \xi_1}\f{\partial}{\partial \xi_2}...\f{\partial}{\partial \xi_n}
                      \log \big\langle e^{\sum_i \xi_i X_i} \big\rangle \Big|_{\xi_i=0},
\eea
where $\langle \cdots \rangle$ means to take the expectation value of the variables. For our purpose, we would take $X_i$ to be the stress energy tensor $T(z_i)$ at point $z_i$, the expectation value to be $\langle... \rangle_\rho$.
We denote the connected correlation functions of $T$ by $U^\rho_n(T(z_1),T(z_2),...,T(z_n))$, and the first few of them are given by
\bea
&&U^\r_1(T(z_1))=\langle T(z_1)\rangle,\nn\\
&&U^\r_2(T(z_1),T(z_2))=\langle T(z_1) T(z_2)\rangle_\r -\langle T(z_1)\rangle_\r \langle T(z_2)\rangle_\r,\nn\\
&&U^\r_3(T(z_1),T(z_2),T(z_3))
= \langle T(z_1)T(z_2)T(z_3)\rangle_\rho- \langle T(z_1) \rangle_\r \langle T(z_2)T(z_3)\rangle_\rho
 -\langle T(z_2) \rangle_\r \langle T(z_1)T(z_3)\rangle_\rho \nn\\
&& \phantom{U^\r_3(T(z_1),T(z_2),T(z_3))=}
 -\langle T(z_3) \rangle_\r \langle T(z_1)T(z_2)\rangle_\rho
 +2\langle T(z_1)\rangle_\r\langle T(z_2)\rangle_\r\langle T(z_3)\rangle_\r.
\eea
We could also generalize to operator $\partial^m T$, for examples,
\bea
&& U^\r_2(\partial T(z_1),\partial T(z_2))
 =\langle \partial T(z_1) \partial T(z_2)\rangle_\r-\langle \partial T(z_1)\rangle_\r \langle \partial T(z_2)\rangle_\r\nn \\
&& U^\r_2(\partial^2 T(z_1), T(z_2))
 =\langle \partial^2 T(z_1)  T(z_2)\rangle_\r-\langle \partial^2 T(z_1)\rangle_\r \langle  T(z_2)\rangle_\r.
\eea
We derive the geometric conditions on the cylinder with coordinate $w$ and spatial period $L$, but now it is convenient to work on complex plane with coordinate $z$. We would like to show that the geometric conditions  is invariant under a conformal map $z=f(w)$.  $S^{(n)}_{A,\rho}(w)\sim \log \langle \sigma_n(w_1)\td{\sigma}_n(w_2)\rangle_{\rho^n}$, under a conformal map $z=f(w)$, we have
\be
S^{(n)}_{A,\rho}(z)\sim \log \langle \sigma_n(z_1)\td{\sigma}_n(z_2)\rangle_{\rho^n} +h_n \log(f'(w_1)f'(w_2)),
\ee
where $h_n=\frac{c}{24}(n-1/n)$. Therefore, the requirement $S^{(n)}_{A,\rho}(w)\sim O(c)$ is equivalent to $S^{(n)}_{A,\rho}(z)\sim O(c)$.
We further use OPE of twist operators on the plane with the coordinate $z$, and get exactly the same conditions for the one-point functions of quasi-primary operators with $\Phi_K(w)$ replaced by $\Phi_K(z)$.
By a conformal map $z=e^{2\pi i w/L}$ the cylinder is mapped to the complex plane.
If the conditions on the plane are justified, it leads to the justification of the conditions on the cylinder.

On the complex plane one may rewrite the first geometric state condition (\ref{result1supp}) as
\be\label{re1}
\f{1}{2\pi i}\oint_{z_2}\f{dz_1}{z_1-z_2} \Big(\lim_{c\to \infty}\frac{U^\r_2(T(z_1),T(z_2))}{c^2}\Big)=0.
\ee
Similarly, for the condition (\ref{result2Dsupp}) we have
\be
\langle \mathcal{D}\rangle_\rho-3 \langle \mathcal{A}\rangle_\rho \langle T\rangle_\rho +2 \langle T\rangle_\rho^3
=\langle(T(TT))\rangle_\rho-3 \langle(TT)\rangle_\r \langle T\rangle_\r+2 \langle T\rangle_\r^3
+\f{9}{10}\big(\langle (\partial^2T T)\rangle_\r-\langle \partial^2 T\rangle_\r\langle T\rangle_\r\big)+O(c),
\ee
and the condition
\be\label{re2}
\f{1}{2\pi i}\oint_{z_3}\f{dz_1}{z_1-z_3} \f{1}{2\pi i}\oint_{z_3}\f{dz_2}{z_2-z_3}
\lim_{c\to \infty}\frac{1}{c^2}\big[ U^\rho_3(T(z_1),T(z_2),T(z_3))+\f{9}{10}U^\r_2(\partial^2 T(z_2), T(z_3))\big]
=0.
\ee
For the condition (\ref{result2Bsupp}), it is given by
\be\label{re3}
\f{1}{2\pi i}\oint_{z_2}\f{dz_1}{z_1-z_2}
\lim_{c\to\infty}\frac{1}{c^2}\big[ U^\r_2(\partial T(z_1), \partial T(z_2))-\f{4}{5} U^\r_2(\partial^2 T(z_1), T(z_2))\big]
=0.
\ee
Higher order conditions (\ref{condI})(\ref{conH})(\ref{conE}) can also be rewritten as the connected correlation functions. We will not show them here.

The geometric state conditions are in analogy to the planar limit of correlation function of large $N$ expansion in four-dimensional Yang-Mills theory in vacuum state. However, there is no solid argument to justify this analogue. If one require a more stronger condition that the connected correlation functions of the scaled operator $T/c$,
\be
U^\rho_n(\partial^{m_1}T(z_1)/c,\partial^{m_2}T(z_2)/c,...,\partial^{m_n}T(z_n)/c)\sim O(1/c^{n-1}),
\ee
for any integer $n$ and $m_1,...,m_n$.
It is just
\be\label{conjecture}
U^\rho_n(\partial^{m_1}T(z_1),\partial^{m_2}T(z_2),...,\partial^{m_n}T(z_n))\sim O(c).
\ee
The conditions (\ref{re1}), (\ref{re2}) and (\ref{re3}) will be satisfied.
One could check the higher order conditions, they all should be satisfied. Note that the conditions we find are a criterion for generic excited states not just for vacuum. We have checked that (\ref{conjecture}) is right for primary state and thermal state up to $n=3$. It begs a quantum gravity interpretation of these conditions.

\section*{Acknowledgments}

We thank Bin Chen, Chong-Sun Chu, Bei-Lok Hu, Wei Li, Wei Song, Chushun Tian and Yong-Shi Wu for helpful discussions.
WZG is supported  by the National Center of Theoretical Science (NCTS).
FLL is supported by Taiwan Ministry of Science and Technology through Grant No.~103-2112-M-003-001-MY3.
JZ is supported in part by Italian Ministero dell'Istruzione, Universit\`a e Ricerca (MIUR), and Istituto Nazionale di Fisica Nucleare (INFN) through the ``Gauge Theories, Strings, Supergravity'' (GSS) research, and by Fondazione Cariplo and Regione Lombardia, Grant No.\ 2015-1253.%

\appendix

\section{Conditions for Geometric CFT states}\label{appA}

The conditions for geometric CFT states are expressed as one-point functions of quasiprimary operators in the vacuum family. We would like to summarize the definitions of these quasiprimary operators up to level $8$, more details can be found in \cite{Chen:2013kpa,He:2017txy}.
At level $2$, we have the quasiprimary operator $T$. At level $4$, we have
\be \label{Adef}
\mA=(TT)-\f{3}{10}\p^2T.
\ee
We use $(\mX\mY)$ to denote normal ordering of $\mX$ and $\mY$, and on the complex plane it is defined as
\be
(\mX\mY)(z)=\f{1}{2\pi i}\oint_z \f{dw}{w-z}\mX(w)\mY(z).
\ee
At level $6$, we have two quasiprimary operators
\bea
&&\mB=(\p T\p T)-\f{4}{5}(\p^2TT)-\f{1}{42}\p^4T,  \nn \\
&&\mD= (T(TT))-\f{9}{10}(\p^2TT)-\f{1}{28}\p^4 T + \f{93}{70c+29} \mc B.
\eea
At level $8$ we have three quasiprimary operators,
\bea\label{defEHI}
&& \mc E= (\p^2 T\p^2 T)-\f{10}{9}(\p^3 T \p T)+\f{10}{63}(\p^4TT)-\f{1}{324}\p^6 T,  \nn\\
&& \mc H= (\p T(\p TT))-\f{4}{5}(\p^2 T(T T))+\f{2}{15}(\p^3 T \p T)-\f{3}{70}(\p^4TT)
        + \frac{9 (140 c+83)}{50 (105 c+11)}\mc E, \nn\\
&& \mc I= (T(T(TT)))-\f{9}{5}(\p^2 T(TT))+\f{3}{10}(\p^3 T \p T)
        + \frac{81 (35 c-51)}{100 (105 c+11)}\mc E
        + \frac{12(465 c-127)}{5 c(210 c+661)-251}\mc H.
\eea

\section{Derivation of geometric conditions}\label{appB}

By requiring the R\'enyi entropy (Eq. (\ref{eq3}) in the main text) to be $O(c)$ we may get the following conditions, i.e., the conditions for geometric states.
With some calculations we get R\'enyi entropy up to $O(\ell^8)$,
\begin{eqnarray}\label{renyi}
&&S^{(n)}_{A,\r}=\frac{c(n+1)}{12n}\log \frac{\ell}{\epsilon}+\frac{\mathcal{C}_2}{1-n} \ell^2+\frac{\mathcal{C}_3}{1-n}\ell^3+\frac{2 \mathcal{C}_4-\mathcal{C}_2^2}{2(1-n)}\ell^4+ \frac{\mathcal{C}_5-\mathcal{C}_2\mathcal{C}_3}{1-n}\ell^5\nn \\
&&\phantom{S^{(n)}_{A,\r}=}+\frac{2 \mathcal{C}_2^3-3 \mathcal{C}_3^2-6 \mathcal{C}_2\mathcal{C}_4+6 \mathcal{C}_6}{6(1-n)}\ell^6\nonumber  +\frac{\mathcal{C}_7+\mathcal{C}_2^2 \mathcal{C}_3-\mathcal{C}_3 \mathcal{C}_4-\mathcal{C}_2 \mathcal{C}_5}{1-n}\ell^7\nn \\
&&\phantom{S^{(n)}_{A,\r}=}+\frac{4 \mathcal{C}_8-4\mathcal{C}_2 \mathcal{C}_6-4 \mathcal{C}_3 \mathcal{C}_5-2\mathcal{C}_4^2+4\mathcal{C}_2^2 \mathcal{C}_4+4\mathcal{C}_2\mathcal{C}_32-\mathcal{C}_2^4}{4(1-n)}\ell^8 +O(\ell^9)
\end{eqnarray}
with
\begin{eqnarray}\label{Cterm}
&&\mathcal{C}_2=b_T \langle T\rangle_\r,\quad \mathcal{C}_3=\frac{b_T}{2} \partial\langle T\rangle_\r,\nn \\
&&\mathcal{C}_4=b_{\mathcal{A}}\langle \mathcal{A}\rangle_\r+b_{TT}\langle T\rangle_\r^2+\frac{3 }{20}b_T \partial^2\langle T\rangle_\r\nonumber \\
&&\mathcal{C}_5=\frac{1}{30} b_T\partial^3 \langle T\rangle_\r
+\frac{1}{2}b_{\mathcal{A}}\partial \langle \mathcal{A}\rangle_\r+\frac{1}{2}b_{TT}\partial\langle T\rangle_\r^2,\nonumber \\
&&\mathcal{C}_6 = b_{\mathcal{B}}\langle \mathcal{B}\rangle_\r+b_{\mathcal{D}}\langle \mathcal{D}\rangle_\r+b_{T\mathcal{A}} \langle T\rangle_\r \langle \mathcal{A}\rangle_\r+b_{TTT} \langle T\rangle_\r^3\nonumber \\
&& \phantom{\mathcal{C}_6 =}
 +b_{\mathcal{K}}  \mathcal{K}_\r+\frac{1}{168} b_T\partial^4 \langle T\rangle_\r+ \frac{5}{36}b_{\mathcal{A}} \partial^2 \langle \mathcal{A}\rangle_\phi +\frac{5}{36}b_{TT} \partial^2 \langle T\rangle_\r^2 ,\nonumber \\
&&\mathcal{C}_7=\frac{1}{1120}b_T \partial^5 \lag T \rag_\r +\frac{1}{36}\big[ b_{\mathcal{A}}\partial^3\lag \mA \rag_\r+b_{TT}\partial^3 (\lag T \rag_\r^2)\big]\nonumber \\
&& \phantom{\mathcal{C}_7 =}
+\frac{1}{2}\big[ b_{\mathcal{B}}\partial\langle \mathcal{B}\rangle_\r+b_{\mathcal{D}}\partial\langle \mathcal{D}\rangle_\r+b_{T\mathcal{A}} \partial(\langle T\rangle_\r \langle \mathcal{A}\rangle_\r)+b_{TTT} \partial(\langle T\rangle_\r^3) +b_{\mathcal{K}} \partial \mathcal{K}_\r\big],\nonumber\\
&& \mathcal{C}_8 = b_{\mathcal{E}}  \langle \mathcal{E}\rangle_\r+b_{\mathcal{H}}\langle \mathcal{H}\rangle_\r+b_{\mathcal{I}}\langle \mathcal{I}\rangle_\r+b_{T\mathcal{B}}\langle T\rangle_\r\langle \mathcal{B}\rangle_\r\nonumber \\
&& \phantom{\mathcal{C}_8 =}+ b_{T\mathcal{D}}\langle T\rangle_\r\langle \mathcal{D}\rangle_\r+b_{\mathcal{A}\mathcal{A}}\langle \mathcal{A}\rangle^2_\r +b_{TT\mathcal{A}}\langle T\rangle^2_\r\langle \mathcal{A}\rangle_\r+b_{TTTT} \langle T\rangle^4_\r\nonumber \\
&& \phantom{\mathcal{C}_8 =} +b_{T\mathcal{K}} \langle T\rangle_\r \mathcal{K}_\r+b_{\mathcal{O}}\mathcal{O}_\r+b_{\mathcal{P}}\mathcal{P}_\r+b_{\mathcal{Q}}\mathcal{Q}_\r
+b_{\mathcal{R}}\mathcal{R}_\r\nonumber \\
&& \phantom{\mathcal{C}_8 =} +\frac{1}{8640}b_T\partial^6 \langle T\rangle_\r+\frac{7}{1584}b_{\mathcal{A}}\partial^4 \langle \mathcal{A}\rangle_\r
+\frac{7}{1584}b_{TT}\partial^4 \langle T\rangle^2_\r\nonumber \\
&& \phantom{\mathcal{C}_8 =} +\frac{7}{52}[ b_{\mathcal{B}}\partial^2 \langle \mathcal{B}\rangle_\r+ b_{\mathcal{D}}\partial^2 \langle \mathcal{D}\rangle_\r+b_{T\mathcal{A}} \partial^2 (\langle T\rangle_\r \langle \mathcal{A}\rangle_\r)\nonumber \\
&& \phantom{\mathcal{C}_8 =} +b_{TTT}\partial^2  \langle T\rangle_\r^3+ b_{\mathcal{K}}\partial^2   \mathcal{K}_\r].
\end{eqnarray}
The expectation values $\lag\mX\rag_\r = \lag\mX(w)\rag_\r$, $\mX=T,\mA,\mB,\mD,\cdots$, are functions of the coordinate $w$.
The coefficients $b_K$ are defined in \cite{Chen:2013kpa} from the OPE coefficients $d_K$ of the twist operators and are constants depending on $n$ and $c$.
There are also definitions
\begin{eqnarray}\label{definitions}
&&\mathcal{K}_\r=(\pET)^2-\frac{4}{5} \ET \ppET, \nonumber \\
&&\mathcal{O}_\r=\pET \pEA -\frac{2}{9}\ET \partial^2 \EA-\frac{4}{5}\ppET \EA,\nonumber \\
&&\mathcal{P}_\r= (\ppET)^2-\frac{10}{9} \pET  \pppET +\frac{10}{63}\ET \partial^4\ET, \nonumber \\
&&\mathcal{Q}_\r=\frac{7}{9}\ET\mathcal{K}_\r,\quad \mathcal{R}_\r=\frac{7}{11}\ET \mathcal{K}_\r.
\end{eqnarray} \\
At $O(\ell^4)$, we have
\be
2 \mathcal{C}_4-\mathcal{C}_2^2
=\frac{n^2-1}{720 n^3}\big[ (n^2-1)(\langle \mathcal{A}\rangle_\r-\langle T\rangle_\r^2)+18 n^2\partial^2\langle T\rangle_\r\big]+O(1/c).
\ee
The last term is $O(c)$, we get the first condition,
\be\label{result1supp}
\lim_{c\to \infty}\frac{\langle \mathcal{A}\rangle_\r -\langle T\rangle^2_\r}{c^2}=0.
\ee
At $O(\ell^5)$, we have
\be
\mathcal{C}_5-\mathcal{C}_2\mathcal{C}_3
=\frac{n^2-1}{2880 n^3} \big[ 5(n^2-1)(\partial \langle \mathcal{A}\rangle_\r-2 \langle T\rangle_\r \partial \langle T\rangle_\r)
+8 n^2 \partial^3 \langle T\rangle_\r\big]+O(1/c).
\ee
This would lead to the condition,
\be
\lim_{c\to \infty} \frac{\partial \langle \mathcal{A}\rangle_\r-2 \langle T\rangle_\r \partial \langle T\rangle_\r}{c^2}=0.
\ee
This is nothing but the derivative of (\ref{result1supp}).
At $O(\ell^6)$, we have
\begin{eqnarray}
&& \hspace{-5mm} \phantom{=}
   2 \mathcal{C}_2^3-3 \mathcal{C}_3^2-6 \mathcal{C}_2\mathcal{C}_4+6 \mathcal{C}_6\nonumber \\
&& \hspace{-5mm} = \frac{n^2-1}{60480 n^5}\Big\{ 35( \langle \mathcal{D}\rangle_\r-3 \langle \mathcal{A}\rangle_\r \langle T\rangle_\r +2 \langle T\rangle_\r^3)
+35\big(\langle \mathcal{B}\rangle_\r - \mathcal{K}_\r-2(\langle \mathcal{D}\rangle_\r-3 \langle \mathcal{A}\rangle_\r \langle T\rangle_\r+2 \langle T\rangle_\r^3)\nonumber \\
&& \hspace{-5mm} \phantom{=} -5[\partial^2 \langle \mathcal{A}\rangle_\r-2 (\partial\langle T\rangle_\r)^2-2 \langle T\rangle_\r\partial^2 \langle T\rangle_\r]\big)n^2
+7\big( 5 \mathcal{K}_\r- 5 (\partial\langle T\rangle_\r)^2+4 \partial\langle T\rangle_\r \partial^2 \langle T\rangle_\r\big)n^3\nonumber \\
&& \hspace{-5mm} \phantom{=}+35\Big[ (\langle \mathcal{B}\rangle_\r- \mathcal{K}_\r)
          -(\langle \mathcal{D}\rangle_\r-3 \langle \mathcal{A}\rangle_\r \langle T\rangle_\r+2 \langle T\rangle_\r^3)
          -5(\partial^2 \langle \mathcal{A}\rangle_\r-2 (\partial\langle T\rangle_\r)^2-2 \langle T\rangle_\r\partial^2 \langle T\rangle_\r)-\frac{36}{7}\partial^4 \langle T\rangle_\r \Big]n^4\nonumber \\
&& \hspace{-5mm} \phantom{=} -7 \big(5 \mathcal{K}_\r- 5 (\partial\langle T\rangle_\r)^2+4 \partial\langle T\rangle_\r\partial^2 \langle T\rangle_\r\big)n^5 \Big\}+O(1/c).
\end{eqnarray}
By using the constraint (\ref{result1supp}), we obtain
\be
\lim_{c\to \infty}\frac{\partial^2 \langle \mathcal{A}\rangle_\r-2 (\partial\langle T\rangle_\r)^2-2 \langle T\rangle_\r\partial^2 \langle T\rangle_\r}{c^2}=0.
\ee
Therefore, we will have the following conditions at this order,
\begin{eqnarray}
&&\label{result2Bsupp}\lim_{c\to \infty}\frac{\langle \mathcal{B}\rangle_\r- \mathcal{K}_\r}{c^2}=0, \\
&&\label{result2Dsupp}\lim_{c\to \infty}\frac{\langle \mathcal{D}\rangle_\r-3 \langle \mathcal{A}\rangle_\r \langle T\rangle_\r +2 \langle T\rangle_\r^3}{c^2}=0.
\end{eqnarray}
The expression of $O(\ell^8)$ is too lengthy, so we just list the results at this order,
\begin{eqnarray}\label{constraint8supp}
&& \label{condI}
   \lim_{c\to \infty}\frac{1}{c^2}\big[\EI-4\ED\ET-3\EA^2
   +12\EA \ET^2+6\ET^4\big]=0,  \\
&& \label{conH}
   \lim_{c\to \infty}\frac{1}{c^2}\Big[45 \EH - 65 \EB \ET + 10 \ET \partial^2\EA
   36 \EA \ppET \nonumber - 72 \ET^2 \ppET \nonumber \\ && \qquad \qquad - 45 \pEA \pET + 90 \ET [\pET]^2\Big]=0,  \\
&& \label{conE}\lim_{c\to \infty}
   \frac{1}{c^2}\big[\EE - [\ppET]^2 - 10/63 (\ET \pfourET
   - 7 \pppET \pET)\big]=0.
\end{eqnarray}
Without loss of generality, we assume the one-point functions $\langle \mX\rangle_\r$  have the following forms,
\begin{eqnarray}
&&\langle T(w)\rangle_\r=\sum_{k=-1}^{+\infty} c^{-k} t_k(w), \quad
  \langle \mathcal{A}(w)\rangle_\r=\sum_{k=-2}^{+\infty} c^{-k} a_k(w), \nonumber\\
&&\langle \mathcal{B}(w)\rangle_\r=\sum_{k=-2}^{+\infty} c^{-k} b_k(w), \quad
  \langle \mathcal{D}(w)\rangle_\r=\sum_{k=-3}^{+\infty} c^{-k} d_k(w), \nonumber\\
&&\langle \mathcal{E}(w)\rangle_\r=\sum_{k=-2}^{+\infty} c^{-k} e_k(w), \quad
  \langle \mathcal{H}(w)\rangle_\r=\sum_{k=-3}^{+\infty} c^{-k} h_k(w), \nn\\
&&\langle \mathcal{I}(w)\rangle_\r=\sum_{k=-4}^{+\infty} c^{-k} i_k(w),
\end{eqnarray}
where $t_k(w)$,$a_k(w)$,$b_k(w)$,$d_k(w)$,$e_k(w)$ and $i_k(w)$ are arbitrary functions of order $O(c^0)$.
The above geometric conditions give some relations among one-point functions $\langle \mX\rangle_\r$.
The result is
\begin{eqnarray} \label{constraintsmath}
&& \ET = c \alpha (w)+\beta (w)+\frac{\gamma (w)}{c}+O\Big(\f{1}{c^2}\Big) , ~~\nn \\
&&   \EA = c^2 \alpha (w)^2+c \delta (w)+\epsilon (w)+O\Big(\frac{1}{c} \Big), \nn\\
&& \EB = c^2 \Big[\alpha'(w)^2-\frac{4}{5} \alpha (w) \alpha''(w)\Big]+c \zeta (w)+O(c^0),\nn\\
&& \ED = c^3 \alpha (w)^3+3 c^2 \alpha (w) [\delta (w)-\alpha (w) \beta (w)]
        +c \eta (w)+O(c^0),\nn\\
&& \EE = c^2 \Big\{\alpha''(w)^2+\frac{10}{63} [\alpha (w) \alpha^{(4)}(w)
                                             -7 \alpha'(w)\alpha^{(3)}(w)]\Big\}+O(c), \nn\\
&& \EH = c^3 \alpha (w)\Big[\alpha'(w)^2-\frac{4}{5} \alpha (w) \alpha''(w)\Big]
        +c^2\Big[- \alpha '(w)^2\beta (w)
                 -2 \alpha (w) \alpha '(w) \beta '(w)
                 + \frac{4}{5} \alpha (w)^2 \beta''(w) \nn \\
&& \phantom{\EH=}
                 + \frac{8}{5} \alpha (w) \alpha ''(w) \beta (w)
                 + \alpha '(w) \delta '(w)
                 - \frac{4}{5} \alpha ''(w)\delta (w)
                 - \frac{2}{9} \alpha (w) \delta ''(w)
                 + \frac{13}{9} \alpha (w) \zeta (w) \Big] +O(c),\nn\\
&& \EI = c^4 \alpha (w)^4
        +2c^3 \alpha (w)^2 [3 \delta (w)-4 \alpha (w) \beta (w)]
        +c^2 [ 12 \alpha (w)^2 \beta (w)^2
              +4 \alpha (w)^3 \gamma (w) \nn \\
&& \phantom{\EI=}
              -12 \alpha (w) \beta (w) \delta (w)
              +3\delta (w)^2
              -6 \alpha (w)^2 \epsilon (w)
              +4 \alpha (w) \eta (w) ] +O(c),
\end{eqnarray}
with $\alpha(w)$, $\beta(w)$, $\gamma(w)$, $\delta(w)$, $\epsilon(w)$, $\zeta(w)$, $\eta(w)$ being arbitrary order $O(c^0)$ holomorphic functions.

\section{A coordinate-dependent example}\label{appC}

Let's consider  a state constructed by superposition of primary state and its global descendants (on the complex plane),
\be\label{globalsuper}
|{\Psi}\rangle:=\mathcal{N}\sum c_m |{\partial^m \phi}\rangle,
\ee
where $\mathcal{N}$ is the normalization constant. For $c_m=\frac{z^m}{m!}$, we could write $|{\Psi_c}\rangle$ as the ``coherent'' state, i.e.,
\be
|{\Psi_c}\rangle=\mathcal{N}e^{zL_{-1}}|\phi\rangle,\quad \text{with}\quad \mathcal{N}=(1-\bar z z)^{h},
\ee
where $h_\phi$ is the conformal dimension of $\phi$. It is obvious $|{\Psi_c}\rangle$ is a local state $O(z)|0\rangle$.
We are interested in the expectation value of $\Phi_K(z)$ in $|{\Psi_c}\rangle$. Generally, we have
\be\label{sum1}
\langle \Psi_c| \Phi_K(x)|\Psi_c\rangle=\mathcal{N}^2 \sum_{s,t}\bar c_s c_t \langle \partial^s\phi|\Phi_K(x)|\partial^t \phi\rangle.
\ee
For $s\ge t$, we have
\be\label{sum2}
\langle\partial^s\phi|\Phi_K(x)|\partial^t\phi\rangle=x^{s-t-h_{\Phi_K}} t!s! \sum_{m\ge s-t}^{s} C_{s-t+m+h_{\Phi_K}-1}^{s-t+m}C_{m+h_{\Phi_K}-1}^{m} C_{2 h_\phi-h_{\Phi_K}+s-m-1}^{s-m},
\ee
while for $s<t$,
\be\label{sum3}
\langle\partial^s\phi|\Phi_K(x)|\partial^t\phi\rangle=x^{s-t-h_{\Phi_K}}t!s!
\sum_{m\ge t-s}^{t} C_{t-s+m+h_{\Phi_K}-1}^{t-s+m}C_{m+h_{\Phi_K}-1}^{m} C_{2 h_\phi-h_{\Phi_K}+t-m-1}^{t-m}.
\ee
From (\ref{sum2}) and (\ref{sum3}) into (\ref{sum1}), we get a simple result,
\be
\langle \Psi_c| \Phi_K(x)|\Psi_c\rangle=C_{\phi\phi \Phi_K}\Big(\frac{z \bar z-1}{(x-z)(1-\bar z x)}\Big)^{h_{\Phi_K}}.
\ee
Using (\ref{sum1})-(\ref{sum3}) we could calculate any state like the form (\ref{globalsuper}) as long as we know the coefficients $c_n$. One could check the one-point functions in the state $|\Psi_c\rangle$ do satisfy all the geometric conditions.
For example, the condition (\ref{conH}) is
\begin{eqnarray}
&&\phantom{=}
  45 \EH - 65 \EB \ET + 10 \ET \partial^2\EA + 36 \EA \ppET\nonumber \\
&&\phantom{=}
  - 72 \ET^2 \ppET - 45 \pEA \pET + 90 \ET [\pET]^2\nonumber \\
&&=\frac{18 c \epsilon _{\phi } \left(1845 c \epsilon _{\phi }-385 c+28\right) \left(z z^*-1\right)^8}{35 (105 c+11) (x-z)^8 \left(x z^*-1\right)^8}\sim O(c),
\end{eqnarray}
where we define $\epsilon_\phi=h_\phi/c$.
But if we slightly change the coefficients $c_m=\frac{z^m}{m!}$, it is very likely the corresponding state will violate the constraints.
At least in this example we can see the geometric conditions we find are highly non-trivial.

\section{ Non-geometric descendant states}\label{appD}

In the main text we show that the primary states would satisfy all the geometric conditions. Like the primary state, descendant states can be viewed as descendant operators inserting on the vacuum. There are infinite descendant states in a Verma module $\mathcal{V}(h,c)$. In this paper we only focus on some special examples that are calculable, for example,  the state $|\psi_1\rangle:=\partial^m\phi(0)|0 \rangle$ and $|\psi_2\rangle:= \partial^{m-2}\tilde{\phi}(0)|0\rangle$, where $\tilde{\phi}:=(T\phi)-\frac{3}{4h+2}\partial^{2}\phi$ is the quasi-primary operator with conformal dimension $h_\phi+2$.

We could calculate the one-point function $\langle T\rangle_{\partial^m \phi}$ and  $\langle \mA\rangle_{\partial^m \phi}$ by using the results  in \cite{Guo:2018pvi},
\begin{eqnarray}
&&\langle T\rangle_{\partial^m\phi}=\frac{\pi ^2 \left[c-24 \left(m+c \epsilon _{\phi}\right)\right]}{6 L^2},\nonumber \\
&&\langle \mathcal{A}\rangle_{\partial^m\phi}=\frac{ \pi ^4}{180 L^4\left(c \epsilon _{\phi }+1\right) \left(2 c \epsilon _{\phi }+1\right)}\big[10 \left(1-24 \epsilon _{\phi }\right){}^2 \epsilon _{\phi }^2 c^4\nonumber \\
&&\phantom{\langle \mathcal{A}\rangle_{\partial^m\phi}=}
   +\epsilon_\phi\big(480 \left(90 m^2+28m+3\right) \epsilon _{\phi }^2-6 (120 m+29) \epsilon _{\phi }+5\big) c^3\nonumber \\
&&\phantom{\langle \mathcal{A}\rangle_{\partial^m\phi}=}
   +\big(480 \left(30 m^3+18 m^2+3 m-1\right) \epsilon _{\phi }-(240 m-22)\big)c\nonumber \\
&&\phantom{\langle \mathcal{A}\rangle_{\partial^m\phi}=}
   +480 m \left(6 m^2-1\right)\big],
\end{eqnarray}
where we define the order $c^0$ constant $\epsilon_\phi= h_\phi/c$.
For the states with heavy descendant that is $m= \td m c$, where $\td m \sim O(c^0)$, we have $h_\phi+m \sim c$ and
\be
\lim_{c\to \infty}\frac{\langle \mathcal{A}\rangle_{\partial^m\phi}-\langle T\rangle^2_{\partial^m\phi}}{c^2}=\frac{8  \tilde{m} \pi ^4 \left(\tilde{m}+\epsilon _{\phi }\right) \left(5 \tilde{m}+8 \epsilon _{\phi }\right)}{L^4 \epsilon _{\phi }}\ne 0.
\ee
Even for $m\sim O(c^0)$ the condition (\ref{condI}) is not satisfied, that is
\be
\lim_{c\to \infty}\frac{1}{c^2}\big(\langle \mathcal{I}\rangle_{\partial^m\phi}-4 \langle \mathcal{D}\rangle_{\partial^m\phi} \langle T\rangle_{\partial^m\phi}+12 \langle \mathcal{A}\rangle_{\partial^m\phi} \langle T\rangle^2_{\partial^m\phi}
\langle \mathcal{A}\rangle^2_{\partial^m\phi}-6\langle T\rangle^4_{\partial^m\phi}\big)
=\frac{6144 \pi ^8 m(m+1) \epsilon_\psi ^2 }{L^8}\ne 0,
\ee
for $m\ne 0$.
For the state $|\psi_2\rangle$ with $m\sim O(c^0)$ we have
\begin{eqnarray}
&&\phantom{=}
\lim_{c\to \infty}\frac{1}{c^2}\big(\langle \mathcal{I}\rangle_{\partial^m \td \phi}-4 \langle \mathcal{D}\rangle_{\partial^m \td \phi} \langle T\rangle_{\partial^m \td \phi}+12 \langle \mathcal{A}\rangle_{\partial^m \td \phi} \langle T\rangle^2_{\partial^m \td \phi} \langle \mathcal{A}\rangle^2_{\partial^m \td \phi}-6\langle T\rangle^4_{\partial^m \td \phi}\big)\nonumber \\
&&=\frac{768 \pi ^8 \left[8 \left(m^2-3m+10\right) \epsilon _{\phi }{}^2+16 \epsilon _{\phi }+1\right]}{L^8}\ne 0.\nonumber
\end{eqnarray}
We will not give the explicit results for state $|\partial^m T\rangle$ and $|\partial^m\mA\rangle$.

\providecommand{\href}[2]{#2}\begingroup\raggedright\endgroup


\end{document}